\documentclass[12pt, a4paper]{article}   	% use "amsart" instead of "article" for AMSLaTeX format

\usepackage[a4paper,margin=3cm,marginparwidth=2cm]{geometry}
\usepackage{graphicx}				% Use pdf, png, jpg, or eps§ with pdflatex; use eps in DVI mode
\usepackage{amsmath,amssymb,amscd}
\usepackage{natbib}
\usepackage{hyperref}
\providecommand{\href}[2]{\texttt{#2}}
\providecommand{\url}[1]{\texttt{#1}}

\numberwithin{equation}{section}

\providecommand{\eqref}[1]{(\ref{#1})}
\newcommand{\eref}[1]{Equation~\ref{#1}}
\newcommand{\fref}[1]{Figure~\ref{#1}}
\newcommand{\Fref}[1]{\fref{#1}} % Use for start of sentence, capitalise Figure word.
\newcommand{\sref}[1]{Section~\ref{#1}}
\newcommand{\aref}[1]{Appendix~\ref{#1}}

\newcommand{\tnote}[1]{}
\newcommand{\enote}[1]{}
\newcommand{\jnote}[1]{}

\newcommand{\Ccom}{C}
\newcommand{\Cmig}{B}
\newcommand{\ccom}{c}
\newcommand{\cmig}{b}

\newcommand{\AdaptedK}{K_\mathrm{200}}

\newcommand{\DKS}{D^\mathrm{(KS)}}

\newcommand{\modelname}{{Two-Trip model}}

\begin{document}

% *****************************************
%
% *** TITLE PAGE
%
% Most journals change this part a lot and I also need other information
% on the preprint version so I tend to construct my own
%

% This is the standard set if commands
%\title{A dynamical analysis of spatial interaction models.}
%\author{James Wilkinson, Theodore Emms, Tim S Evans\thanks{Use for Address}}
%\date{\today}							% Activate to display a given date or no date
%\maketitle

%%% \begin{flushright}
%%% %\textbf{Confidential}. \\ Not to be redistributed without permission of the authors.\\
%%% \texttt{Imperial/TP/18/TSE/2} \\
%%% %\eprint{cond-mat/yymmnnn} \\
%%% %23rd March 1998, final corrections 8th April 1999 \\
%%% %\tsecompldate \\
%%% %(\texttt{\jobname.tex}  LaTeX-ed on \today ) \\
%%% %{next preprint number}\\
%%% \end{flushright}
%%% \vspace*{0.5cm}

\begin{center}
{\Large\textbf{Dynamical Analysis of Spatial Interaction  Models}\tnote{tnotes such as this not present in final version}  } \\[0.5cm]
 {\large {James Wilkinson}, {Theodore Emms},
  \href{http://www.imperial.ac.uk/people/t.evans}{Tim S.\ Evans}}
 \\[0.5cm]
\href{https://www.imperial.ac.uk/complexity-science}{Centre for Complexity Science}, and \href{http://www3.imperial.ac.uk/theoreticalphysics}{Theoretical Physics Group},
Imperial College London, SW7 2AZ, U.K.
 \\[0.5cm]  30th August 2018
%  \\[0.5cm]  Keywords: Spatial Interaction Models, Gravity Model, Radiation Model, Zipf's law, Urban Geography
\end{center}

\begin{abstract}
We develop a novel dynamical method to examine spatial interaction models (SIMs). For each SIM, we use our dynamical framework to model emigration patterns. We look at the resulting population distributions to see if they are realistic or not. We use the US census data from 2010 and various spatial statistics to access the success or failure of each model.  While we looked at over eighty different SIMs, we will focus here on two examples: the production constrained gravity model and the Radiation model. The results suggest that all these models fail to produce realistic population distributions and we identify the flaws within existing models. This leads us to suggest that we should define site attractiveness in terms of a second short range SIM leading to a new spatial interaction model --- the {\modelname} --- which offers significant improvements when examined via our method. We also note that our Two-Trip adaptation can be used in any spatial modelling contexts, not just emigration.
\end{abstract}

% *** END of TITLE PAGE
%
% *****************************************

% **********************************************************************************
\section{Introduction}

Spatial interaction models (SIMs) predict flows that occur between a set of sites distributed in a space \citep{openshaw1976empirical}. They have been used to describe population distributions \citep{zipf1946p,B64a}, traffic flows \citep{JWS08}, the movements of cargo ships \citep{KKGB10}, the flow of goods and money \citep{evans2009using}, telephone call patterns \citep{LBKHPSD08}  and friendship networks \citep{lee2011investigating}. Perhaps the most important application of SIMs lies in modelling human movement, where they enjoy an extensive range of applications from city planning \citep{batty2013new,B16e} to archaeological modelling \citep{rihll1987spatial,knappett2008modelling,PB16a,ER16}.

%\citep{rihll1987spatial,knappett2008modelling,evans2009using,PB16a,ER16,BCCH17}.

However the vast majority of studies are static. That is they use a model where the flows between sites do not vary over time so the system is in equilibrium.
The central idea in this paper is to interpret the flows modelled by a SIM literally.  In the context of migration, these flows tell us how many people are moving from one site to another. It is therefore entirely logical to use this information to tell us how the populations at different sites evolve. Including such dynamical feedback within a SIM has been done before \citep{HW78,W08,WD11a,WD11} but most studies using a SIM focus on flows within the system at one particular time.

%\tnote{Methods for testing the accuracy of SIMs have typically involved comparisons between predicted and observed migration. These comparisons to migration data have been shown to produce potentially misleading results \citep{knudsen1986matrix, baxter1983estimation, wilson2011entropy}.}

Our approach is very flexible and can be applied to many different SIM.  We tried over eighty different models though for brevity we focus here primarily on the production constrained Gravity model \citep{wilson1967statistical} and the Radiation model \citep{simini2012universal}. For comparisons with a realistic spatial distribution we used the location of population centres in the USA. One of our conclusions is that no existing SIM was found to be satisfactory by our testing methodology.  Looking at how traditional models fail to match the observed population distributions led us to produce a novel ``{\modelname}'' to illustrate what is missing in traditional models.  We found this {\modelname} with just one additional parameter produced excellent results.

% ***************************************************************************
\section{Methodology}

Our approach has four stages. First we choose a SIM. Second we generate an artificial set of sites whose positions will remain fixed. Next, we use a dynamical extension of a static SIM to evolve the site populations. Lastly, we take various statistical measures of the resulting pattern of populations, stopping the evolution at the point where we get the best match with a reference data set.

% ------------------------------------------------------------------------------
\subsection{Generic Model Definition}
\label{ModelSection}

There are a plethora of SIMs available making it difficult to select the right model for a problem. Part of this stems from ambiguities in the interpretations of model parameters. We will work with a standard form for a SIM, which will also illustrate the ambiguities and choices that need to be made. Consider the typical form of many of these models
\begin{equation}
    T_{ij} = A_{ij} O_i D_j f_{ij}
    \label{gravitygen}
\end{equation}
where we define the following quantities:
\begin{description}
    \setlength{\itemsep}{0pt}
    \setlength{\parskip}{0pt}
    \setlength{\parsep}{0pt}
\item[$T_{ij}$]  the population flow from site $i$ to site $j$,
\item[$A_{ij}$]  a normalisation term, depending on the constraints of the model,
\item[$O_i$]  a source term with a larger value encouraging more trips to originate from site $i$,
\item[$D_j$]  an attractiveness term with a larger value encouraging more trips to the site $j$,
\item[$f_{ij}$]  the deterrence function which determines the cost scale of an individual movement from $i$ to $j$. Invariably this is a function which decreases as the distance between sites increases.
\end{description}
Many elements of \eref{gravitygen} are not precisely defined so choices have to be made, often based on some mixture of expert opinion and computational or data driven necessity.
For example, in the unconstrained Gravity model, first developed in 1781 \citep{monge1781memoire}, it is typical to use the  population of sites, denoted here by $Q_i$  for site $i$, to represent both the source and attractiveness terms: $O_i=D_i=Q_i$. There is no constraint in this simple model so $A_{ij}=1$. The definition of the deterrence function is usually  ambiguous.  In some cases it is linked to some cost function $c_{ij}=-\ln(f_{ij})$ via entropy maximisation arguments \citep{wilson1967statistical} (see also \cite{ES90} and \aref{app:commutingderivation}) but effective costs, be they social, financial, political and so forth, are typically unknown. A common candidate, perhaps driven by a mixture of data availability and expert opinion, is to set the cost proportional to the distance.  When cost is chosen that way and when entropy is optimised subject to a constraint on total cost, this leads to $f_{ij}=\exp(-\beta r_{ij})$ where $\beta$ is a model parameter while $r_{ij}$ is the distance between two sites. Even then, there are many ways to characterise the perceived distance between two points \citep{ERK11}. This includes straight-line distance, the shortest distance by road and the average travel time to name just a few. Distance can also be measured by the rank of the distance to nearbouring sites, i.e.\ the intervening opportunities measure \citep{S40}. So there are many choices to be made when using SIMs. Later in \sref{sresults} we will use a couple of models to illustrate our key findings and we will specify the choices made for the SIMs at that point.

%The choices made when defining SIMs are well illustrated by the following two examples we focus on in this paper: the production constrained Gravity model \citep{wilson1967statistical} and the Radiation model \citep{simini2012universal}.

% --------------------------------------------------------------
\subsection{Dynamical Extensions of Spatial Interaction Models}

%The central idea in this paper is to interpret the flows modelled by a SIM literally.  In the context of migration these flows tell us how many people are moving from one site to another. It is therefore entirely logical to use this information to tell us how the populations at different sites evolve.
%Using SIMs in a dynamical context has been done before \citep{HW78,W08,wilson2008phase,WD11} but most studies using a SIM focus on flows within the system at one particular time.
%the vast majority of these studies focus on flows within the system at one particular time.\tnote{``one particular time''.  Really? What does this mean?} \jnote{``but such such studies have refrained from examining spatial interaction models, as we intend to, and focused instead on their application.''}

The focus of our work is the interpretation of the flows defined by a SIM in terms of changes in populations over time.
However, there is no fixed prescription for how populations of centres change in response given the flow matrix $T_{ij}$ defined by a SIM. So we use the simplest and most natural approach in which the rate of population of site $i$, $Q_i$, changes at a rate that is proportional to the net population movement in and out of that site. That is we conjecture that
\begin{equation}
    \frac{d Q_i}{dt} = \lambda \sum_{k} (T_{ki} - T_{ik})
    \label{eq:dynsim}
\end{equation}
where the flows $T_{ki}$ and $T_{ik}$ and the populations $Q_i$ are those at time $t$. Here $\lambda$ is a rate that can be interpreted as the fraction of a flow $T_{ki}$ who move permanently to a site $k$ from site $i$ per unit time.\footnote{There is an important issue here: machine time vs.\ physical time. Equations like \eqref{eq:dynsim} are often used as a numerical technique to find the solution for a static SIM i.e.\ to find the flows where the non-linear input/output constraints are satisfied given the input/output site sizes. In that case the `time' variable is just a machine time with no physical meaning. We, on the other hand, want to think of the $t$ variable as a physical time.}

In principle we can insert any static SIM into this equation provided we can identify the source terms $O_i$ and the attractiveness terms $D_i$ of \eqref{gravitygen} in terms of our dynamical population variable $Q_i$.
%We will use the same identifications as those used in the corresponding static model.

Once the population flows are defined in terms of populations at time $t$ from a given SIM, this differential equation \eqref{eq:dynsim} was then solved numerically using the Euler method to evolve an initial population up to a final time $\tau$. It was found, however, that models do not reach a steady state. Therefore, this time-scale $\tau$ is treated as an additional parameter of the model and found by optimisation, see \sref{sresults}.

%The procedure used to identify the optimal parameters of the models is discussed in \sref{sresults}.

% % ...............................................................
% \subsubsection{Static SIMs}

% We will be working with models of the generic form given in \eqref{gravitygen}.  However as noted above there many variations of the general form used in practice.

% The deterrence function $f_{ij}$ is a function of distance where a larger distance represents a larger notional cost of making a particular trip.  The precise nature of this cost, financial, social, political or a mixture of many factors, is rarely specified a measured.  The effect of such costs on travel is also rarely known precisely. So in practice the deterrence function is usually a simple function of distance which becomes smaller as the distance between two sites becomes larger. In this paper we will work with Gravity models where the deterrence function is
% \begin{equation}
% f_{ij} = \exp (-\beta r_{ij})
% \end{equation}
% where $\beta$ is a model parameter while  $r_{ij}$ is the actual distance between two sites which are fixed properties of the system.

% A further complication is that there are many ways to specify the distance between two points such as the length of the direct path between two sites or the time of travel. For this paper we will just be using the former. The one exception is the Raidation model which we discuss below in more detail.

% ------------------------------------------------------------------------------
\subsection{Site Generation}
\label{MapSection}

Models must be run on a selection of settlements, or \textit{maps}, with  spatial and population information. Generating maps by sampling data would mean at time $t=0$ the system would match the data perfectly. This could introduce false positive results, so randomly generated maps were used. This approach also allows us to estimate the uncertainties in our results.

Here 1000 settlements were distributed uniformly on a square of 1000 by 1000 spatial units each with a population of 100,000.  In these units the density is $\rho =10^{-3}$ giving a typical site separation scale of $\rho^{-1/2} \approx 30$. The scales used for population and distance are arbitrary and so the absolute values are of little significance.

% ...............................................................
\subsection{Statistic Selection}

An important choice is that of the statistics used when comparing the output of our models to real world data.
%A number of observables and statistics can be identified that overcome the issues with static comparisons and have been shown to be consistent in real world systems. \tnote{Citation to back up claim that ``have been shown to be consistent in real world systems''?}
We work with two statistics that have proved particularly useful in our study.

% ......................................................................
\subsubsection{Population-rank distributions} \label{sec:populationrank}

%Add in the zipf.pdf image somewhere
Of particular interest and significance to this paper is the relationship between city populations and their rank within a system. This is invariably a fat-tailed distribution, often approximated as a power-law, a property formally stated as Zipf's law \citep{Z49,rosen1980size}. Zipf's law has been used in discussions of city size-rank distributions for over a century across virtually all systems for which there are data, for example see \citep{Z49,B64a,rosen1980size}.

%This highlights the source of our interest in the population-rank distribution: owing to its robustness and universality, it represents an ideal comparison to the real world.

We are not concerned with the precise details of the shape of these city size vs.\ rank distributions, whether they are power laws or follow some other particular form.  Our interest is that the shape of these fat-tailed distributions are fairly similar and so this is a robust feature of population dynamics. We will therefore demand that an acceptable dynamical SIM must be able to reproduce this type of distribution. This type of prerequisite for any serious growth model has been suggested before \citep{gabaix1999zipf}. One result of our work is to note that achieving a reasonable match between our dynamical models of the form \eqref{eq:dynsim} and this Zipfian type behaviour is surprisingly difficult to achieve.

To be more precise, we will be comparing the output from our dynamical SIMs to census data from the United States of America \citep{UScensusdata} using the two sample Kolmogorov-Smirnov test \citep{massey1951kolmogorov} (KS test). The KS test quantifies the similarity of two arbitrary curves through the statistic $\DKS$, which is between zero and one but which approaches zero as the two curves become more similar. In order to make this comparison, it is necessary to reduce the United States census data to a 1000 settlement representation. For this reason, 1000 settlements were randomly sampled within the census data and averaged over 100 iterations. This practice yields both a mean and standard-deviation range of the US population-rank distribution.

% .............................................................
\subsubsection{Adapted Ripley's $K$ Statistic}

A key feature of real world systems is the spatial distribution of settlements, this information is not encoded by the population-rank distribution. In order to address this issue, an adaptation of the Ripley's $K$ function \citep{dixon2002ripley} was selected as a suitable secondary statistic.

The Ripley's $K$ function \citep{dixon2002ripley} is widely used to detect deviations of a distribution of points from spatial homogeneity for varying distance scales \citep{bevan2006multiscalar}, see \aref{app:RipleyK} for a definition and discussion.
% For perfect spatial homogeneity, Ripley's K function is $K(r) = \pi r^2$ while clustering of sites causes the functions to increase in value. This behaviour is illustrated in \fref{fig:ripleyseg} %is this necessary?
However, the Ripley's K statistic is a point statistic.  That is the statistic is uninfluenced by the populations of each site and only depends on their presence.
In our case, the sites have both a spatial location and a size.  If we applied the traditional Ripley's K measure to data from any of our models, we would end up with a results close to the minimum possible, $K(r) = \pi r^2$, indicating a homogeneous distribution, since we choose our site locations at random.

However we are particularly interested in the larger sites. In order to capture both site size and site location in one statistic, we adapt Ripley's K statistic and use the statistic $\AdaptedK(r)$ where
%\begin{equation}
% \AdaptedK (r) = \frac{1}{200} \sum_{i \neq j | i,j \in \mathcal{Q}(200)} %\theta(r-r_{ij})
% \label{AdaptedK}
%\end{equation}
\begin{equation}
 \AdaptedK (r) = \frac{1}{200}
 \sum_{i \neq j} \theta(r-r_{ij}) \, \theta(Q_i-Q_\mathrm{200}) \, \theta(Q_j-Q_\mathrm{200})
 \label{AdaptedK}
\end{equation}
where $i$ and $j$ run over the indices of the largest 200 settlements of our 1000 so $Q_\mathrm{200}$ is a little smaller than the population of the site ranked 200-th by size.
%\enote{I changed this because the limits of the sum were very big and complicated, which created a lot of ugly white space - The old version is commented out}
%where $\mathcal{Q}(200)$ is the set of indices of the 200 largest sites.
Spatial boundaries effect this statistic and we outline how we deal with edge effects in \aref{app:RipleyK}.

%Talk about the quantitative measure of fit used (Chi squared or KS test). A qualitative analysis was found to be enough...?

% -----------------------------------------------------------------------
\subsection{Numerical Issues}

% \emph{I think a numerics subsection in the methods is needed. Some idea of the time taken to run each model, methods needed to search for optimal solutions, libraries and language used.  Were there any clever tricks needed?}

% Some of these are mentioned elsewhere
% \begin{itemize}
% \item Euler method for evolution of ODE, step size and $\lambda$ values. (J)
% \item Finding optimised value of parameters. Method used? (J)
% \item Finding intervening opportunities at each step? (T)
% \item General time issues. Python vs. C. (T)
% \end{itemize}

\subsubsection{Stability}

Whilst the $\lambda$ parameter in \eqref{eq:dynsim} can be interpreted as the fraction of people who move permanently to a site, it is also directly linked to the time scale within the system. Therefore, the parameter is of interest from the perspective of stability.

A stability analysis was carried out over a $T=1000$ time-step simulation, using deviations in total population size as a measure of error. Defining the `amplification factor' $\eta$ as the ratio of this population deviation over the first half of the simulation to the final half, so that
\begin{equation}
\eta = \frac{P(0;T/2) - P_0}{P(T/2;T) - P_0}
\end{equation}
where $P(t_1;t_2)$ is the average total population between times $t_1$ and $t_2$ and $P_0$ is the initial total population. A necessary criteria for stability is the amplification factor to be less than unity.

%My version:
%Alternative methods to evaluate stability were also used to verify this result,  all of which yielded similar results. In order to ensure stability of the method we chose to use $\lambda = 0.01$.

It was found that stability demanded values of $\lambda$ as low as 0.1. Alternative methods to the one proposed here were also used, all of which yielded similar results. In order to be sure of our method's stability, we choose to use $\lambda = 0.01$.

%is zero sum? I have a proof of what you're trying to say that I can copy/paste in if you want that might make what you're saying clearer? Also, might be too verbose. You could just say $\frac{dQ}{dt} = \frac{d \left( \sum_{i}Q_i \right) }{dt} = \sum_{i} \frac{dQ_i}{dt} = 0$
% The cause of this instability is unclear given that the residual flow method outlined in \eqref{eq:dynsim} is zero-sum thus making total population fluctuations theoretically impossible. However, we hypothesise that it could be caused by rounding error associated with the 64-bit floats used, compounded by the non-linear effects of the models over many successive iterations.

% ...............................................................
\subsubsection{Programming Issues}

We tested around eighty different models in the course of our research. To achieve this we structured the code carefully to ensure that each model could be implemented   with minimal changes to existing code, with many models implemented in minutes which allowed for real time discussions.

The limiting factor in our work was  model running time. We used standard desktop computer technology and generic interpreted Python code as our run times were typically just 20 seconds for models that could benefit from pre-calculation optimisations. However, some models could not benefit from these optimisations and had much longer running times, some theoretically taking several millennia to reach completion. In these cases, in order for models to run in a reasonable amount of time, we then implemented these models in C++. Using compiled C++ with customised code optimisations produced programmes with far superior performance compared to our Python code implementations but at the cost of dramatically increased development time.
%For example, implementing the Radiation model in C++ took approximately 100 hours.

\subsubsection{Evaluating Intervening Opportunities}

One of the primary difficulties when building an implementation of the Radiation model \cite{simini2012universal} is in making it efficient enough to run in a practical time frame. This problem arises here as the number of intervening opportunities between two sites \cite{S40} must be recalculated at every point in time since the sites sizes (the `opportunities') are varying.
Consider, naively, the number of calculations that must be performed in determining the intervening opportunities for a single pair of sites $s_{ij}$ within the system. Here, one must test whether any of the  $N$ settlements fall within radius $r_{ij}$: a test which takes $N$ comparisons and must be performed $N^2$ times to calculate every element of $T_{ij}$. The result is a time complexity of $\mathcal{O}(N^3)$.
There are a few tools that a programmer can reach for to reduce the running time, such as pre-calculating the intervening settlements. However, this turns a $\mathcal{O}(N^3)$ time complexity into a $\mathcal{O}(N^3)$ memory complexity that would require terabytes of RAM to run on the US dataset. Neither of these solutions is feasible and so a balance must be struck between memory and CPU usage. The implementation used in this study was developed over an extended period of time and resulted in a $\mathcal{O}(N^2)$ memory complexity and a $\mathcal{O}(N^2 \log N )$ time complexity.

% **************************************************************************
%\section{Spatial Interaction Models}
\section{Results}
\label{sresults}

So far we have defined a dynamical extension to any SIM through \eqref{eq:dynsim}. Once we have a form for the flows $T_{ij}$ we can investigate the output of different models and compare this against the data from the US census.
The values chosen for any model parameters, including the running time $\tau$, were those which gave a population-rank distribution which most closely resembled that of US Census data, where the Kolmogorov-Smirnov test statistic was used to measure closeness of fit.

We will start to specify our models by discussing a generic framework which applies to most SIMs. In our case we tested approximately eighty existing models \citep{E17,W17}, most being some type of Gravity model. However, we will summarise our findings by considering two exemplary cases: the production constrained Gravity model \cite{wilson2011entropy} and the Radiation model \citep{simini2012universal}, which we define later in this section.

% ------------------------------------------------------------------
\subsection{The Production Constrained Gravity Model}

% We begin by analysing the production constrained Gravity model, interpreting population as both the source and attractiveness terms $Q_i=D_i=O_i$ and choosing a deterrence function of the form $f_{ij} = \exp(-\beta r_{ij})$ \citep{wilson2011entropy}. This model has some real world applications, for instance \citep{HT05}, but it may be too simple for many problems. The flow of population between two sites, from $i$ to $j$, in the Production Constrained Gravity model is given as
% \begin{equation}
%     T_{ij} = A_i Q_i Q_j e^{-\beta r_{ij}}
%     \label{eq:gravprodconstr}
% \end{equation}
% where $r_{ij}$ is the Euclidean distance between sites. We impose the production constraint that the number of people leaving each site is equal to the the population of that site, $Q_i = \sum_j T_{ij}$. This results in the following form for $A_i$
% \begin{equation}
%     A_i = \left[ \sum_{k} Q_k e^{-\beta r_{ik}} \right]^{-1} %added this in
%     \, .
% \end{equation}

We begin by analysing the production constrained Gravity model, interpreting population as both the source and attractiveness terms $Q_i=D_i=O_i$ and choosing a deterrence function of the form $f_{ij} = \exp(-\beta r_{ij})$ \citep{wilson2011entropy} where $r_{ij}$ is the Euclidean distance between sites. This model has some real world applications, for instance \citep{HT05}, but it may be too simple for many problems. We also impose the production constraint that the number of people leaving each site is equal to the the population of that site, $Q_i = \sum_j T_{ij}$. This results in the following form the flow of population $T_{ij}$ from site $i$ to site $j$,
\begin{equation}
    T_{ij} = A_i Q_i Q_j e^{-\beta r_{ij}}
    \, ,
    \label{eq:gravprodconstr}
    \qquad
    A_i = \left[ \sum_{k} Q_k e^{-\beta r_{ik}} \right]^{-1} %added this in
    \, .
\end{equation}

\begin{figure}[htb]
  \begin{center}
      \includegraphics[width=0.9\textwidth]{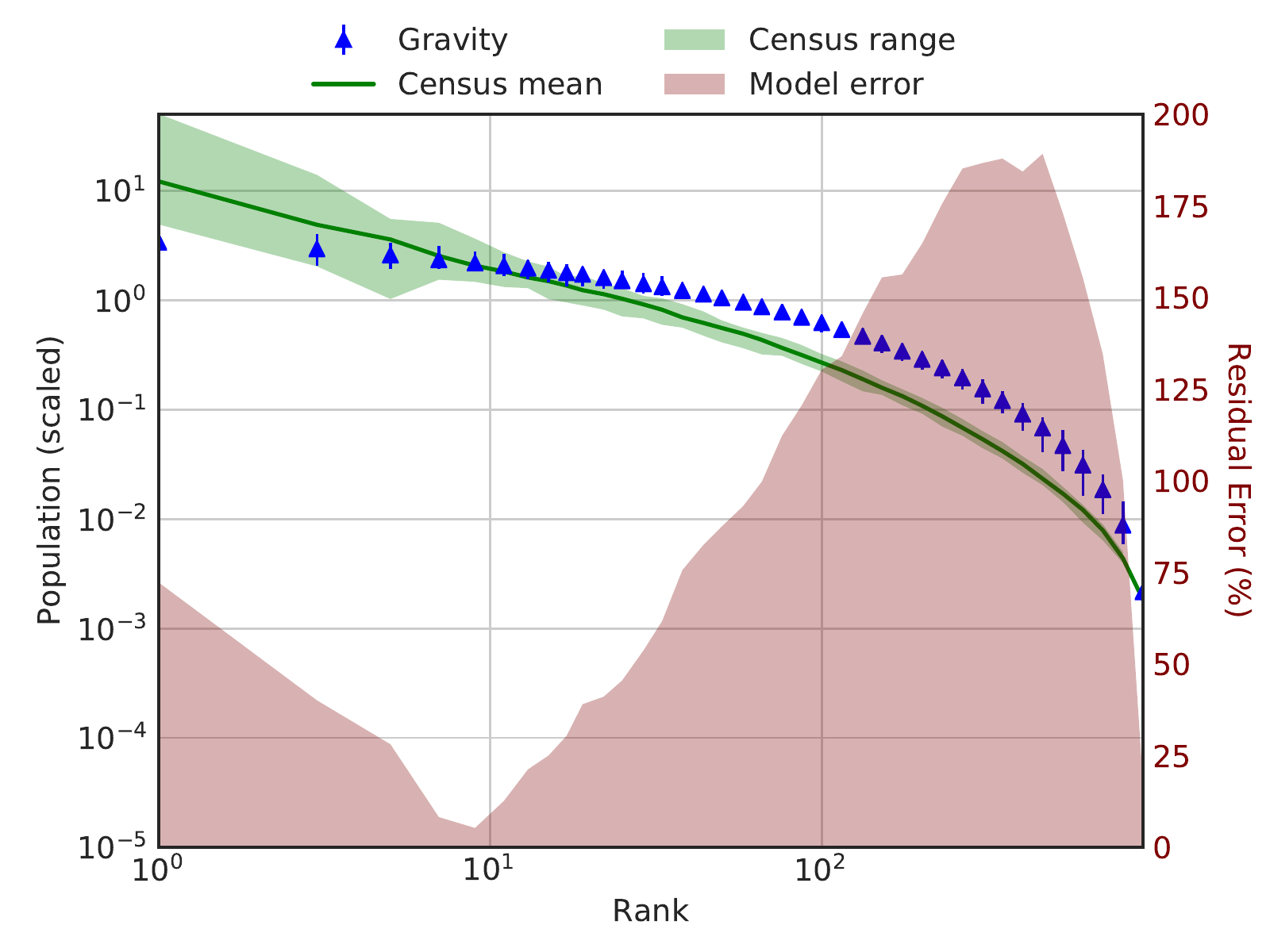}
  \end{center}
      \caption{The optimal population-rank distribution for \eqref{eq:gravprodconstr} with a KS test statistic of $\DKS=0.133$. Model parameters were $\beta = 0.023$, $\lambda = 0.01$ and $\tau = 1000$. The model's population rank distribution (blue triangles) are plotted against sampled US census data (green line) and the data range generated from the sampling process (green fill). Each data point is representative of a settlement within the model's system, averaged over 100 runs, each on a different random map. Error bars and census range displayed are three interquartile ranges. Percentage residual between the model and mean data is also displayed (red fill).}
      \label{wilsonzipf}
\end{figure}

The optimal population rank distribution generated by the Production Constrained Gravity model, is displayed in Figure \ref{wilsonzipf}.
The small error bars, produced as three interquartile ranges on 100 differing systems, displayed in \fref{wilsonzipf} demonstrate that the rank distribution produced by the model is insensitive to the underlying system. This is a quality that is observed in the real world.
Given our measure of the intrinsic variation in our results, it is clear that the population-rank distribution given in \fref{wilsonzipf} is unrealistic. The largest cites in the Production Constrained Gravity model are too small while the smallest cities are too large. Clearly a more successful model needs to have a larger flow into the bigger settlements.

\begin{figure}[!h]
  \begin{center}
      \includegraphics[width=0.9\textwidth]{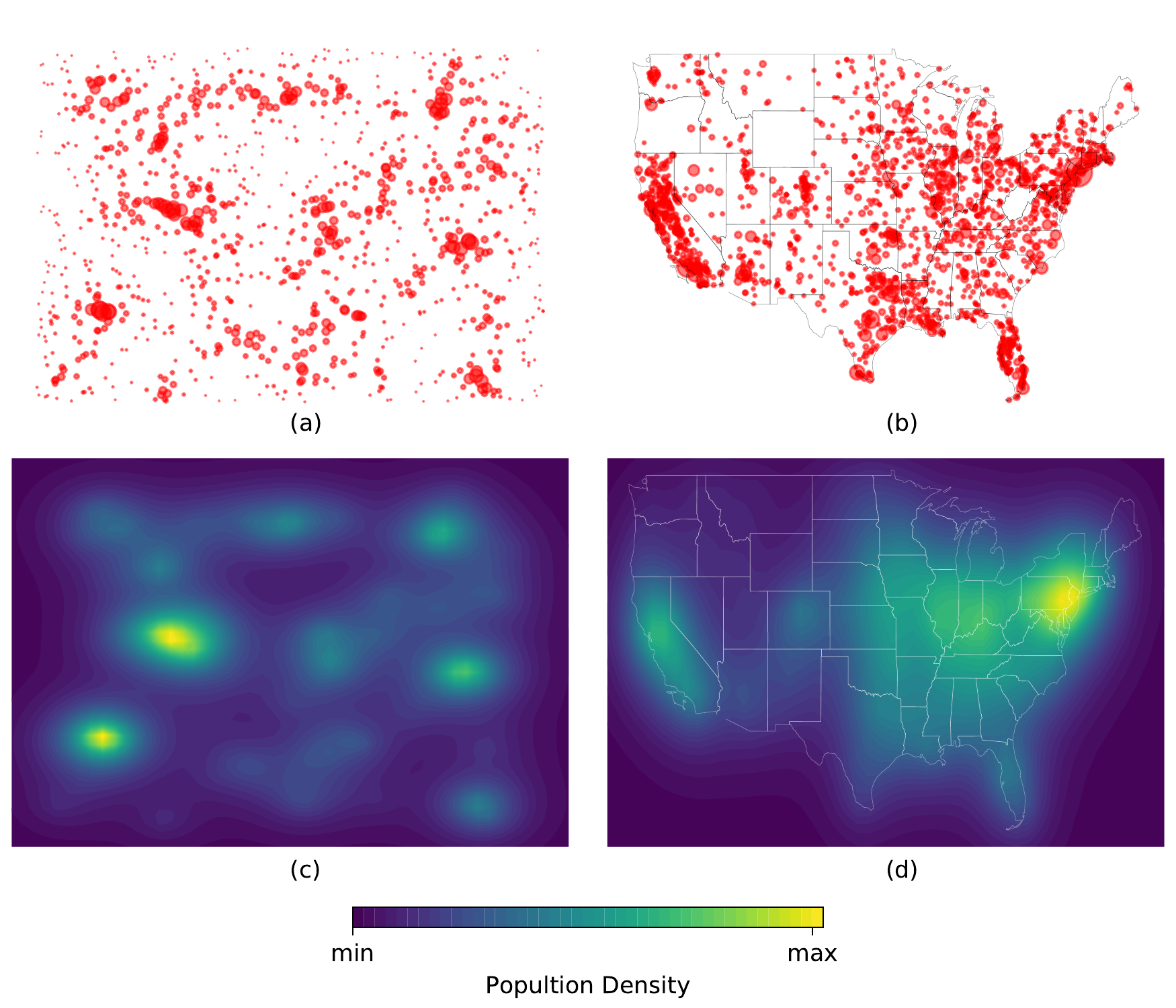}
  \end{center}
      \caption{A typical output of a dynamic SIM, displaying settlement locations and sizes (a) and a population density map (c), produced in this case by the production constrained Gravity model, run via our dynamic method for $\tau = 1000$ time steps with parameters $\lambda = 0.01$ and $\beta = 0.023$. The results of both plots are contrasted to the United States census data \citep{USshapefile, UScensusdata} in (b) and (d). The radius of the red circles is proportional to their population. It can be seen that the system's evolution results in clusters of settlements not unlike that seen exhibited in the United States.}
      \label{exampleoutput}
\end{figure}

There are a plethora of models offering a suitable extension to the gravity model. For example, one used to study the location of retail outlets \citep{H64,LH65,HW78,rihll1987spatial,RW91} redefines attractiveness as a power of the destination's population, so $D_j = {Q_j}^{1 + \epsilon}$ where $\epsilon$ is a fixed parameter. This would have the effect of giving the larger cities more ``pulling power'' with regards to population flow, and should increase the size of the largest cities. However, we found that this model also fails to recover the observed population rank distribution. Approximately eighty such model extensions were tested with mostly negative results \citep{E17,W17}.  This does not mean that these alterations were not worth investigation; many of them shed light onto the nature of an effective model.

\subsection{The Radiation Model}

The form of the Radiation model is similar to that of the Production Constrained Gravity Model. The major difference is that the Radiation model \citep{simini2012universal} uses Stauffer's ``intervening opportunities'' \citep{S40} $s_{ij}$ as measure of distance and not the geographical distance $r_{ij}$ we have used in other models. To define this measure precisely we define the sets $C_{ij}$ containing the indices of all settlements closer to settlement $i$ than settlement $j$, i.e.\ $C_{ij} = \{ k | r_{ik}<r_{ij}\}$. The number of intervening opportunities $s_{ij}$ between site $i$ and $j$ is then
\begin{equation}
 s_{ij} = \sum_{ k  \in C_{ij} } D_k  \, ,
 \label{intopp}
\end{equation}
where $r_{ij}$ is the distance between site $i$ and site $j$. The intervening opportunities measure $s_{ij}$ can be interpreted as another measure of the distance from site $i$ to site $j$.

We again consider the case $D_k=Q_k$ where a site's attractiveness is equal to its population.
%\tnote{Do we need this next bit?  `-- has the side effect of considering not only a site's population but also the population in the surrounding area.' Maybe.}
The original derivation of the Radiation model \citep{simini2012universal} used a simple probabilistic model which exploits record statistics \citep{SJ13}. It was based on the idea that people will consider the closest opportunities before those further away, regardless of size of the actual physical distance $r_{ij}$. The form of the Radiation model we use gives the flows $T_{ij}$ from site $i$ to site $j$ as
\begin{equation}
    T_{ij}
    =
    Q_i \frac{Q_i Q_j}{\left(Q_i + s_{ij} \right) \left( Q_i + Q_j + s_{ij} \right)} \, .
    \label{radmodel}
\end{equation}
Note that in a finite size system the Radiation model in this form is only approximately an output constrained model, $\sum_j {T_{ij} } \approx Q_i$ but the approximation is very good here in practice.

This form of the Radiation model is also parameterless in the sense that it has no global model parameter which requires fitting to data; there are no free parameters such as the $\beta$ in our production constrained Gravity model \eqref{eq:gravprodconstr}. The only effective parameter in our dynamical form of the Radiation model is the time $\tau$ the model is run for.

On the other hand, the biggest drawback numerically in using the Radiation model is that the intervening opportunities measure of distance is continually changing in our dynamical model. In the static case, the way the intervening opportunities measure $s_{ij}$ adapts to different distance scales present in the geography is put forward as one of its advantages. However with changing $Q_i$ here the computational workload of recalculating the  intervening opportunities measure $s_{ij}$  becomes the limiting factor in our numerical simulations of a dynamical Radiation model.

\begin{figure}[htb]
  \begin{center}
      \includegraphics[width=0.9\textwidth]{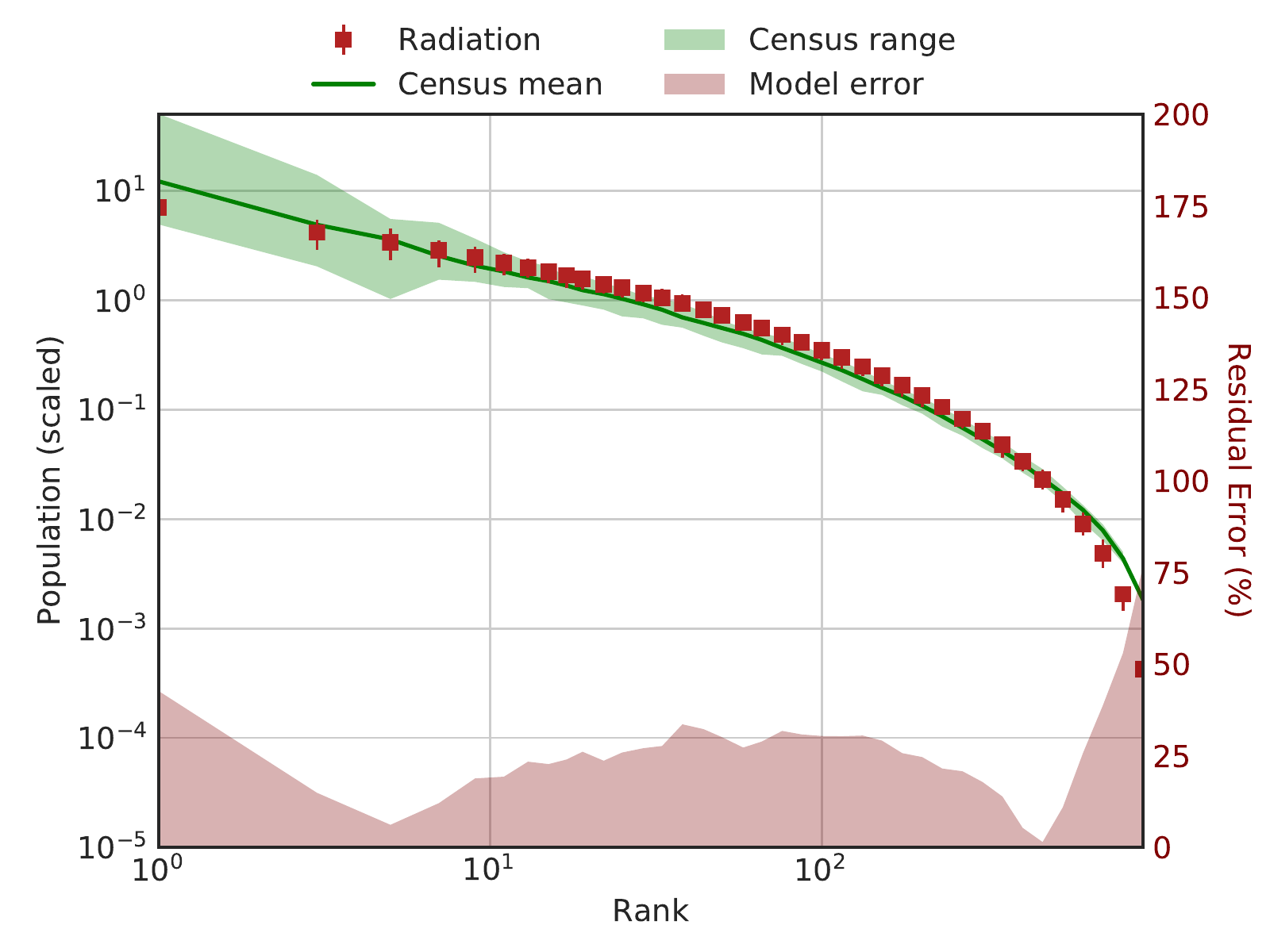}
  \end{center}
      \caption{The optimal Zipf distribution for the Radiation model. The optimal KS-Test value is $\DKS=0.053$. The model parameters used were $\lambda = 0.01$ and $\tau = 1040$. The model's population rank distribution (red squares) are plotted against sampled US census data (green line) and the data range generated from the sampling process (green fill). Each data point is representative of a settlement within the model's system, averaged over 100 runs. Error bars and census range displayed are three interquartile range. Percentage residual between the model and mean data is also displayed (red fill).}
      \label{radiationzipf}
\end{figure}

The Population-Rank distribution in \fref{radiationzipf} seems to show improvement over the Production Constrained Gravity model and many  other Gravity models investigated. Still, the Radiation model does not agree with the census data for small populations; roughly speaking the smallest seventy percent of the settlements fall outside the bounds of error. On the other hand, the higher quality fit of the Radiation model to the US Census data suggests that the Radiation model captures some  dynamics missing from other models.

\begin{figure}[!t]
  \begin{center}
      \includegraphics[width=0.9\textwidth]{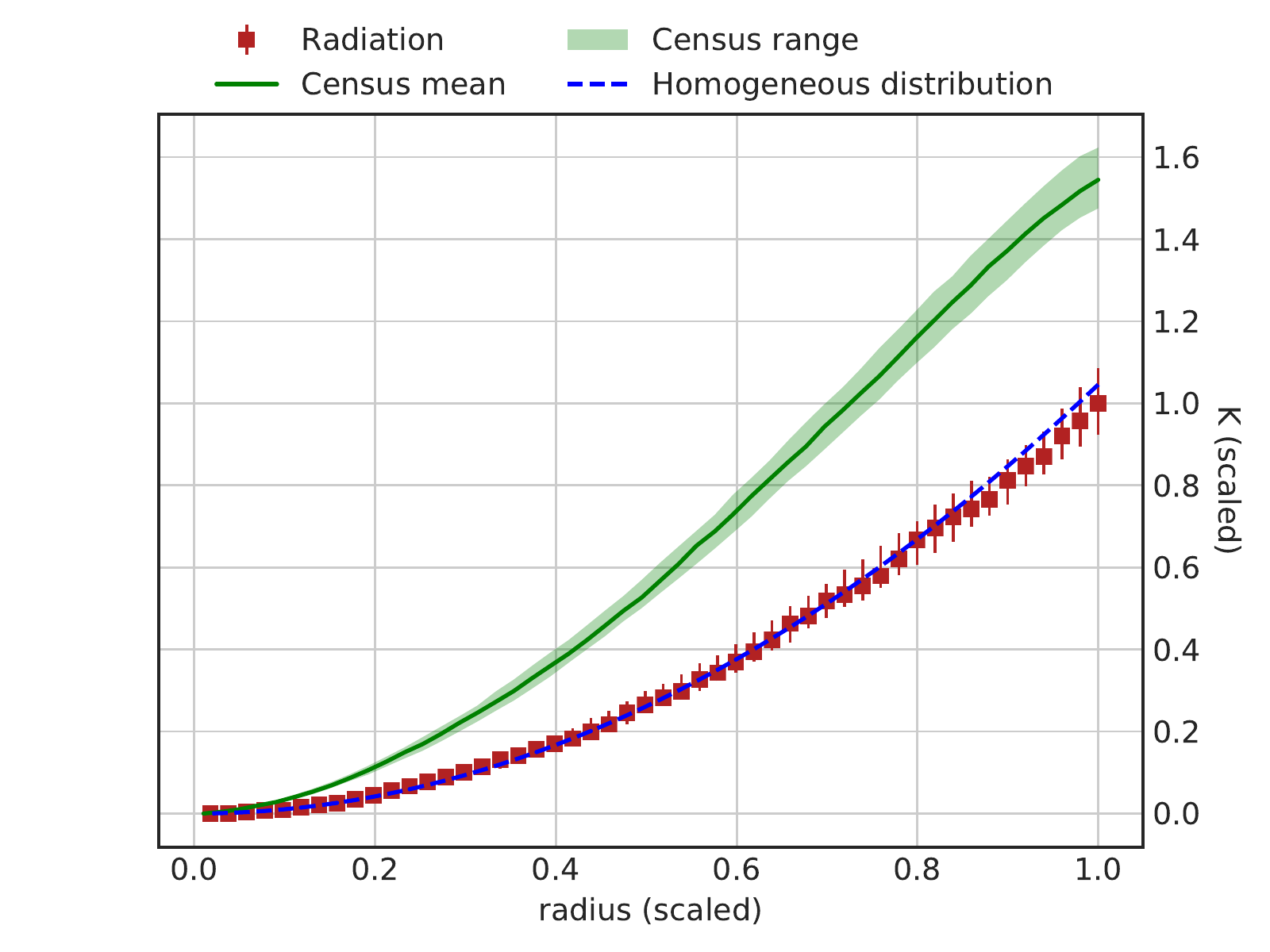}
  \end{center}
      \caption{Optimal results for our adapted Ripley's K statistic, $\AdaptedK(r)$ of \eqref{AdaptedK}, for the Radiation model with parameters $\lambda = 0.01$ and $\tau=1040$. Error bars are interquartile ranges both on the model's Ripley's K function, and also on the census data. The dashed-blue homogeneity line represents the expected Ripley's K function for a uniformly distributed set of points, following the form of $\hat{K}(r) = \pi r^2$.}
      \label{radiationrk}
\end{figure}

Unfortunately, an analysis of the spatial distribution of large settlements in the Radiation model, as shown in \fref{radiationrk}, indicates a potential flaw in the model. The Radiation model produces spatial distributions that appear to be completely random, reflecting our initial conditions. We expect that in reality large settlements often cluster together, as seen in the US census data. In terms of our adapted Ripley's K statistic, $\AdaptedK(r)$ of \eqref{AdaptedK}, other models produce distributions that resemble data more closely than the Radiation model so we interpret this result as a flaw in this dynamical version of the Radiation model.

%Talk about performance? Maybe worth a very brief mention as it does make the model less useful.

%Another issue with the Radiation model is its $ \mathcal{O}\left( n^3 \right) $ time complexity in the number of settlements when implemented naively. This model required a sophisticated implementation in order to be runnable in small enough amount of RAM in a reasonable amount of time. Early implementations of the model would have demanded several tens of gigabytes of RAM and centuries of running time to evaluate. The final versions of this algorithm ran in less than one hundred megabytes of RAM and took approximately a week of single threaded CPU time.

% -------------------------------------------------------------------------------
\subsection{The {\modelname}} \label{sec:movenear}

Looking at the results from the eighty or so SIMs we have studied, as illustrated by the two main examples discussed above, we arrive at four key observations:
\begin{enumerate}
	\item Models aware of population in the neighbourhood of destination settlements outperform other models.
    \item Gravity models \eqref{eq:gravprodconstr} require alterations that encourage movement into large settlements.
    \item In all existing models, places with no population at any time have zero attractiveness and they never grow. This is not desirable, as in reality all settlements have started with zero population.
    \item The attractiveness changes suddenly at the boundary of settlements. The boundaries of sites often reflect political rather than actual divisions and so in many cases they should not have such a dramatic effect on the dynamics of a region.
\end{enumerate}
As we could not find an existing model which performed well on our two measures or which addressed our key observations, we set out to construct a new model which addressed these issues and to see if it would perform better.

%This model works by treating the destination of a migration as a waypoint on a longer journey

\begin{figure}[htb]
\centering
\includegraphics[width=0.9\textwidth]{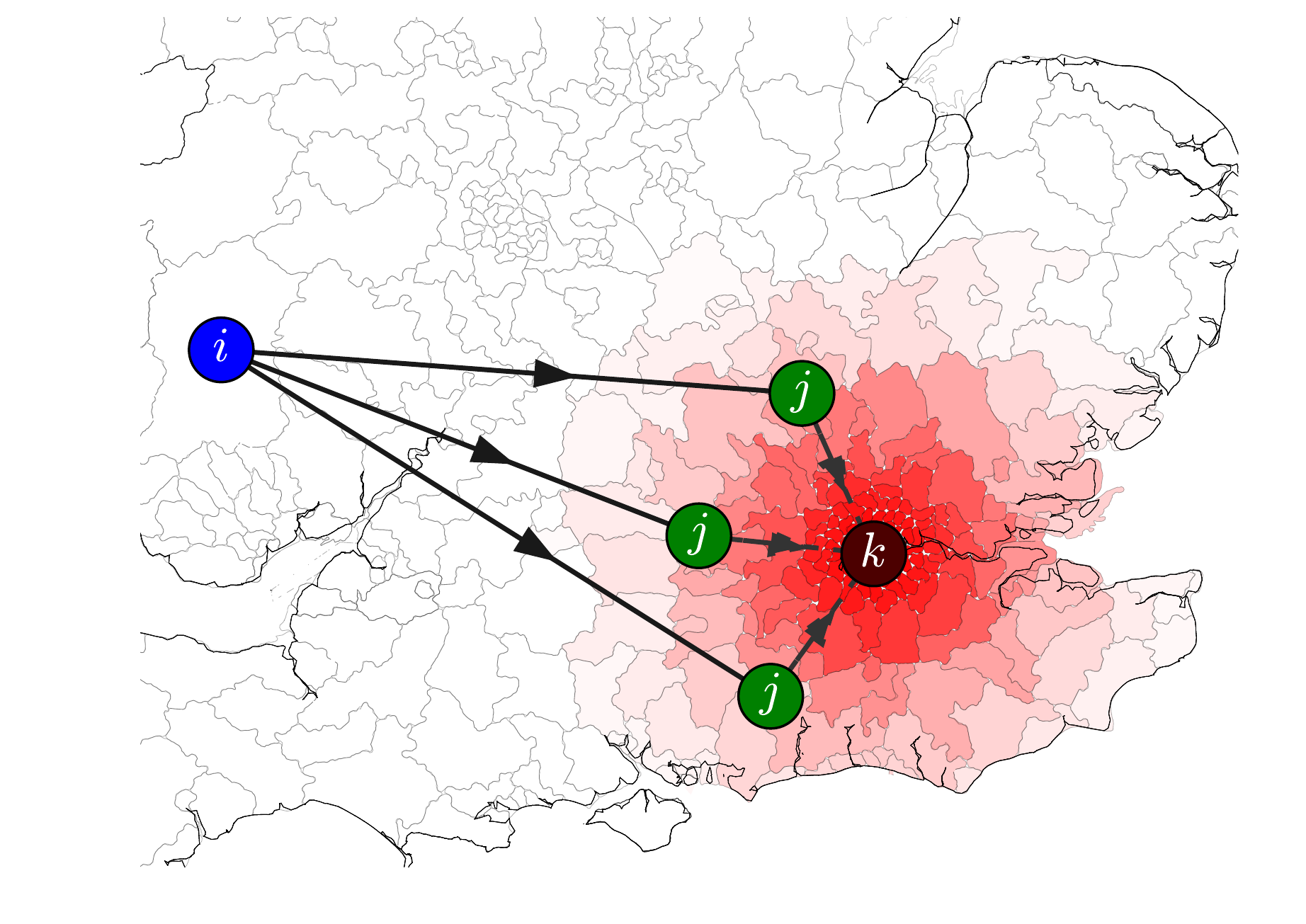}
\caption{The commuting model considers pairs of trips, $i \to j \leftrightarrow k$.  The first part $i \rightarrow j$ represents occasional movements occurring on long time scales, emigration, while the second trip captures temporary, short-distance exchanges, commuting trips, between sites $j$ and neighbouring sites $k$. When a large site $k$ is in the proximity of $j$, this augments the attractiveness of $j$ as commutes are more ``convenient''. In this case we imagine a migration from rural Wales, site $i$, to possible sites $j$, satellite communities which lie within daily commute of central London, site $k$.}
\label{fig:commutingdemo}
\end{figure}

We propose the \textit{{\modelname}} in which the attractiveness of a settlement takes account of the population in a region close to that settlement. The attractiveness of a site $j$ is not just based on local opportunities, but also on opportunities accessible through short-distance ``commutes'' to nearby sites.
In our migration context, the model now considers a trip between three locations: $i \to j \leftrightarrow k$. The trip from $i \to j$ is the conventional permanent migratory movement while the trip $j \leftrightarrow  k$ is temporary, happening on a short time scale and not altering any settlement's population.
This concept is illustrated in figure \ref{fig:commutingdemo}.

We implement this concept in the simplest possible way, defining the attractiveness $D_j$ of a site $j$ in terms of a new parameter $\gamma$ and the commuting flow from $j$ to nearby sites as modelling in a simple Gravity model:
\begin{equation}
	D_{j} = \sum_k Q_k e^{-\gamma r_{jk}} \, .
    \label{twostepattr}
\end{equation}
We imagine that attractiveness is only enhanced by nearby sites so we will assume that $\gamma \gg \beta$.
Here site $j$ is included in the sum so that its population makes a contribution to the attractiveness in the usual way, enhanced by the populations of close neighbours  so that, $D_{j} = Q_j + \sum_{k| k \neq j} Q_k e^{-\gamma r_{jk}}$.

This form for the attractiveness clearly has the first desired property in that it includes the effect of neighbouring settlements. This form of attractiveness will encourage growth of settlements surrounding large cities --- a feature absent from prior models. The resulting, highly attractive settlement clusters provide increased migration from small to large settlements, fulfilling the second property that was desired from a model. It also solves our third key observation as now a site can grow even if its population is zero. This definition also gives a continuous attractiveness at every point in space. The discontinuities present in other models have been removed. This reduces the impact of the arbitrarily chosen boundary of settlements on the dynamics of the system, addressing the fourth key observation.
This interpretation of the attractiveness \eqref{twostepattr} is also appealing as the distance of one's commute to work is an important consideration when relocating. This reformulation of attractiveness would make locations with short commutes to large cities very attractive. All four key observations outlined in the introduction to this section are therefore addressed by this form for attractiveness.

%We also expect that this will allow settlements surrounding large cities to benefit from their large neighbour, and so regions of large settlements will form, presumably on a length scale set by $\gamma^{-1}$.
%These clusters of large settlements should be very attractive, and so we also expect that the population will be encouraged to move from small settlements to larger ones. This is the second property that was desired in the model.
%Additionally, these large, attractive settlement clusters encourage increased migration from small to large settlements, thus fulfilling the second property that was desired in the model.

%I like this point but I don't know where it belongs

%It is useful to recognise that a settlement is an arbitrarily selected area of land on which people live with a boundary that does not exist in a physical sense. If the model treats the attractiveness of a settlement as a function of the population of that settlement alone, then the model has "hard", physical boundaries. It makes sense to impose no boundaries: a pseudo-settlement created by grouping settlements should behave in the same way as a single, large settlement. We can test to see if models satisfy this condition analytically. This has been something done by academics in the past and used to criticise models \citep{evans-conversation}. However, human beings are aware of these boundaries so they could exist in a “soft”, social sense. Unless it is demonstrated that sociological boundaries do not exist in reality, this criticism may be invalid.

In principle we can use our enhanced attractiveness \eqref{twostepattr} in any SIM with the limiting case $\gamma \rightarrow \infty$ in \eqref{twostepattr} giving us the usual attractiveness of a site $j$ in terms of population of that site $\lim_{\gamma \rightarrow \infty} D_j=Q_j$.  Even models using the intervening opportunities measure of distance \eqref{intopp} can use the {{\modelname}}  attractiveness of \eqref{twostepattr} to represent the number of opportunities in a site in place of the population.

In our study we chose to adapt the production constrained Gravity model using the attractiveness given by \eqref{twostepattr}.  The full form for our {{\modelname}} is then
\begin{eqnarray}
 T_{ij}
 &=&
 A_i Q_i \left( \sum_k Q_k e^{-\gamma r_{ik}} \right) e^{-\beta r_{ij}} \, ,
 \label{twostepmodel}
 \\
 A_i
 &=&
 \left[ \sum_{j} \left( \sum_k Q_k e^{-\gamma r_{ik}} \right) e^{-\beta r_{ij}} \right]^{-1} \, .
\end{eqnarray}
A more formal derivation of this model from an entropy viewpoint is given in  \aref{app:commutingderivation}.
The {{\modelname}} has two parameters, which can both be interpreted as distance scales: $(1/\beta)$ gives the distance scale over which changes of residence occur, and $(1/\gamma)$ sets a short term (`commuting') movement distance scale.

\begin{figure}[htb]
  \begin{center}
      \includegraphics[width=0.9\textwidth]{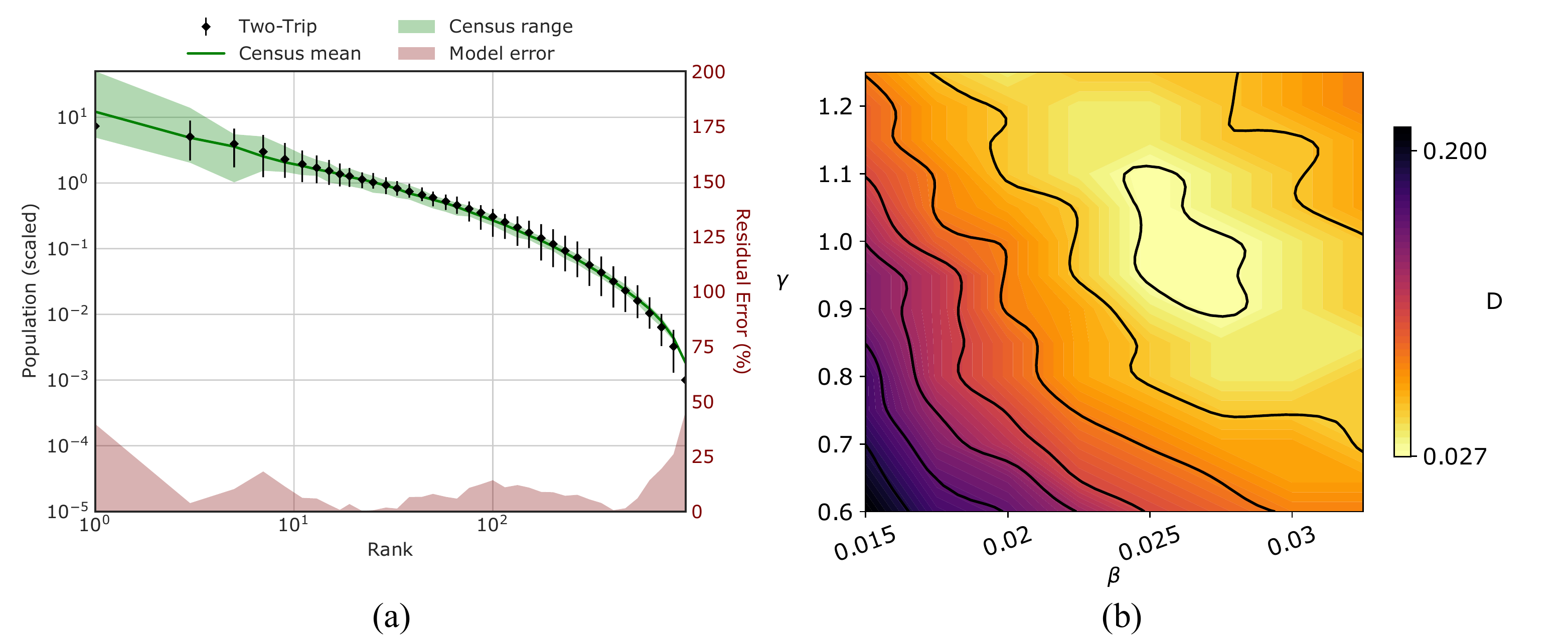}
  \end{center}
      \caption{The Population-Rank distribution for the {\modelname}. This model achieved a remarkably low KS test score of $\DKS=0.0267$, the lowest value seen in this study. The model parameters used were $\beta = 0.025$, $\gamma = 1$, $\lambda = 0.01$ and $\tau = 1000$. The model's population rank distribution (black diamonds) are plotted against sampled US census data (green line) and the data range generated from the sampling process (green fill). Each data point is representative of a settlement within the model's system, averaged over 100 runs. Error bars and census range displayed are a third quartile range. Percentage residual between the model and mean data is also displayed (red fill).}
      \label{movenearzipf}
\end{figure}

\begin{figure}[htb]
  \begin{center}
      \includegraphics[width=0.9\textwidth]{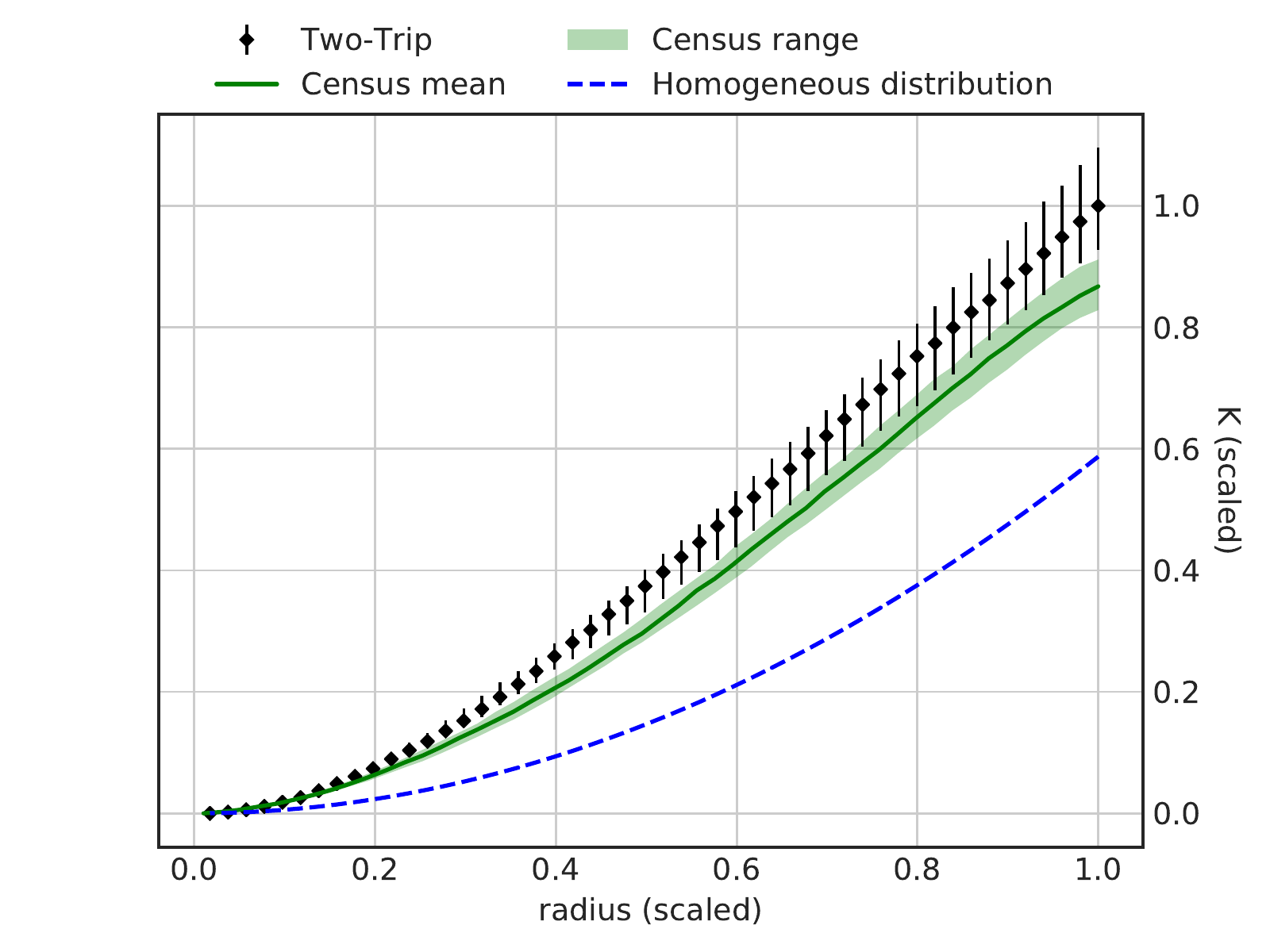}
  \end{center}
      \caption{Optimal results for our adapted Ripley's K statistic, $\AdaptedK(r)$ of \eqref{AdaptedK}, for the {\modelname} with parameters $\lambda = 0.01$, $\beta = 0.025$, $\gamma=1.0$ and $\tau=1000$. Error bars are quartile ranges both on the model's Ripley's K function, and also on the census data. The dashed-blue homogeneity line represents the expected Ripley's K function for a uniformly distributed set of points, following the form of $\hat{K}(r) = \pi r^2$.}
      \label{movenearrk}
\end{figure}

\Fref{movenearzipf} exhibits the population-rank distribution of the {{\modelname}} model, demonstrating a faithful reproduction of the distribution generated from census data, outperforming all other models in this investigation including the Radiation model. We also note that the best fit to the data is obtained with a model where $\beta\ll \gamma$ so that the emigration length scale $\beta^{-1}=40.0$ is forty times longer than the commuting length scale $\gamma^{-1}=1.0$.

At first sight, these distance scales seem too small given that in our units the average distance between sites is roughly $30$.  The exponential form of our emigration and commuting deterrence functions in \eqref{eq:gravprodconstr} and \eqref{twostepattr} then tells us that an average site has essentially no commuter satellite sites and its population will only emigrate to a small set of its immediate neighbours. However, the {{\modelname}} plays the exponential fall off in distance in these deterrence functions against the slower fall off with area that the Poisson distribution shows for the probability of one site having a neighbour within a certain radius.  That is we can estimate that the closest neighbour that any one site of our $N=1000$ sites has is at distance $r_1$ which is roughly\footnote{To arrive at this estimate we demand that the probability $p_1$ that we have one site within radius $r_1$ to be $1/(N-1)$. This is so that on average we expect only \emph{one} site of all $N=1000$ sites present will have such a close neighbour at a distance $r_1$.  The sites are Poisson distributed so $p_n = \exp(-\lambda) \lambda^n/n!$ is the probability of finding $n$ sites within an area $A$ with $\lambda=\rho A$. With $\rho =10^{-3}$ we have that $r_1 \sim 0.40$.  Thus we see that the commuting distance scale $\gamma^{-1}$ we have found in our optimal {{\modelname}} is on the scale of the closest nearest neighbour distance $r_1$ in our model.} $0.40$.  What this means is that only a handful of sites will have enough close neighbours for the commuting process to generate an attractiveness $D_i$ which is significantly greater than their population, $Q_i$. That in turn means that the {\modelname} is boosting the size of a few sites where they are in spatially-tight clusters. Because of the spatial statistics in our model, there will only be a few such clusters so we will only boost the population of a few regions, giving us the enhancement of a few larger sites that we needed to match the data Zipf plots and at the same time giving us the clustering of large sites that we see in the data and as measured by our $\AdaptedK(r)$ statistic.
This means this {\modelname} model also outperforms the Radiation model with regards to spatial distribution of large settlements, as is apparent in  \fref{movenearrk}.

The $\chi^2$ statistic for the {\modelname} when compared with data is $308$, which was the lowest seen in any model studied during this project by a large margin. In order to make a meaningful comparison: the Radiation model produced a $\chi^2$ statistic equal to $14000$ which is clearly a much less faithful fit to data.

%The {\modelname} model also outperforms the Radiation model with regards to spatial distribution of large settlements, as is apparent in Figure \ref{radiationrk}. The $\chi^2$ statistic for the {\modelname} when compared with data is $443$, which was the lowest seen in any model studied during this project by a large margin. In order to make a meaningful comparison: the Radiation model produced a $\chi^2$ statistic equal to $9750$ which is clearly a much less faithful fit to data.

% ************************************************************
\section{Conclusion}

% At the outset of this study, a host of problems with the practices found in the literature were identified, centering around limited testing methods and potentially misleading analytical practices \citep{knudsen1986matrix, baxter1983estimation, wilson2011entropy}.
% These issues stem from the treatment of SIMs in a static context, that is the populations are considered at a fixed point in time. However, if the models describe movements of people over longer time-scales, then SIMs have to be realised as part of a dynamical framework in which the populations are changing in response to migration. By adopting this approach, many existing SIMs have been shown to be inadequate for use over longer time-scales.

We started our work from the observation that SIMs (Spatial Interaction Models) are often used to describe  movements of people where the sizes of centres of population remain unchanged. On longer time scales this is not consistent and so such static SIM models ought to be considered in a dynamical context. We have considered one possible dynamical framework, specified in \eqref{eq:dynsim}, and we have used this to make comparisons between many different models using the population-rank and spatial distributions of settlements as a measure of success.

  Over the course of this study, over eighty different existing SIMs were implemented and tested. However, for clarity in this paper we have focussed on two popular examples: the production constrained Gravity model \citep{wilson1967statistical} and the Radiation model \citep{simini2012universal}. These two models had different strengths and weaknesses, but both illustrated why we arrived at our four key observations where improvements were necessary. Firstly a basic Gravity model requires alteration to the cost or attractiveness terms that encourages large settlement growth. Secondly that models which account for the population neighbouring a site outperform those that do not. Finally sites with zero population must be able to re-emerge.

In response to this, we developed a new type of attractiveness $D_j$ of \eqref{twostepattr} and illustrated this with a novel model of spatial interaction, the {\modelname}. In our definition, a site's attractiveness depends not only its size, but on the size of nearby settlements. Our interpretation is that we are considering two journeys between three locations: a short-distance, short term ``commute'' along with a permanent (or at least long-term) migration. Our {\modelname} can be derived formally using an Entropic viewpoint as we show in \aref{app:commutingderivation}. Our {\modelname} was shown to produce a vastly improved population-rank distribution over other models examined in this study. In addition, the resulting spatial distribution of large sites was seen to be a drastic improvement over that produced by the Radiation model.

Our Two-Trip adaptation may also be of use outside the emigration-commute context which we have used here to frame our discussion. In many situations where spatial interactions are of interest, the attraction of any one site will be a function, in part, of the neighbouring sites. The attractiveness of a large centre such as New York, London or Shanghai will be in part because of the large number of neighbouring centres that generate additional resources. The interactions possible at one central site such as London are much larger than the population of central London because many people are prepared to make a short trip in from a satellite community and this adds value to that central site. In our work, we have modelled this enhanced attractiveness effect using a simple Gravity model containing one extra parameter, our short distance scale $\gamma^{-1}$, but any SIM could be used for this second shorter trip.  Indeed the SIM used for the long distance and short distance effects need not be the same depending on the context of the problem. This enhancement of a site's attractiveness caused by its neighbours leads to additional non-linear effects. This reminds us of models such as the retail gravity model \citep{H64,LH65,HW78} where the attractiveness of a site is modelled as a power of the population, $D_j=(Q_j)^\beta$, as used in a dynamical context in \citep{WD11}. Our Two-Trip framework gives an alternative microscopic approach to such non-linear enhancements of the attractiveness of a site.

\section*{Acknowledgements}

TSE would like to thank Ray Rivers for useful discussions.

% *****************************************************************
%
% BIBLIOGRAPHY
%
% **************************************************************

% ---------------------------------------------------------------
% This version uses BibTeX.  Either point it to the dagshared.bib file
% which you will have to update as needed.  Otherwise point it to your own
% bibtex file.

%\newpage
%\bibliographystyle{unsrt}
%\bibliographystyle{abbrv}
%\bibliographystyle{plain}
%\bibliographystyle{acm}
%\bibliographystyle{siam}

%\bibliographystyle{ieeetr}
%\bibliographystyle{apsrev}
%\bibliographystyle{nar}

% Harvard Styles
% see https://www.overleaf.com/18715284tmdmdjncnzcw#/70420455/
%\bibliographystyle{plainnat} % works with _ in URL
%\bibliographystyle{unsrtnat} % works with _ in URL, but unsorted
%\bibliographystyle{abbrvnat} % works with _ in URL
%\bibliographystyle{agsm} % fails with _ in URL
%\bibliographystyle{dcu} % fails with _ in URL

%\bibliographystyle{naturemag-doi}

% Local File
%\bibliography{jamesbib,theobib}
%\bibliography{dynamicalsim}

%\IfFileExists{/Users/time/OneDrive/PAPERS/JabRef/TSEjabrefdatabase.bib}{\bibliography{/Users/time/OneDrive/PAPERS/JabRef/TSEjabrefdatabase}}{}

\newpage

% APPENDIX for dynamical SIM

% ************************************************************

\appendix
\renewcommand{\thesection}{\Alph{section}}
\setcounter{section}{0}

\section{The {\modelname}}\label{app:commutingderivation}

We follow the approach of \citet{wilson1967statistical} (or see, for example, \citet{ES90}). We begin by considering the number of possible micro-states possible for a given set of two-stage population flows through a system of sites
\begin{equation}
W = \frac{T!}{\prod_{i j k \alpha} T_{ijk \alpha}!} .
\end{equation}
Here $\alpha$ indexes each possible example of the route connecting sites $i$, $j$ and $k$, i.e.\ in the language of statistical physics $\alpha$ labels the microstates for each route. The total number of trips made in the system given as $T=\sum_{ijk\alpha} T_{ijk\alpha}$.
The entropy of this system is then
\begin{equation}
 S = \ln(W) = \ln\bigg(\frac{T!}{\prod_{ijk\alpha} T_{ijk \alpha}!}\bigg)  \, .
\end{equation}
We now assume that all properties of the system are identical for each microstate $\alpha$ of one route $ijk$. So we may assume approximately similar flows across each of the inter-personal connections $\alpha$ and hence that $T_{ijk\alpha} \approx {T_{ijk}}/{N_{ijk}}$ where $N_{ijk}$ is the number of these identical connections possible along the $ijk$ route (the number of microstates in the coarse grained state $ijk$). With this simplification and using Sterling's approximation we find that
\begin{equation}
 S \approx \ln(T!) - \sum_{ijk} \left( T_{ijk} \ln\bigg(\frac{T_{ijk}}{N_{ijk}}\bigg) - T_{ijk} \right)
\end{equation}
We aim to maximise this system entropy under a set of constraints, which can be summarised as a production constraint,
\begin{equation}
\sum_{jk} T_{ijk} = O_i \, ,
\label{outconstr}
\end{equation}
a total migratory (first stage) cost constraint
\begin{equation}
\sum_{ijk} T_{ijk} \cmig_{ij} = \Cmig \, ,
\label{migconstr}
\end{equation}
and a commuting (second stage) cost constraint
\begin{equation}
\sum_{ijk} T_{ijk} \ccom_{jk} = \Ccom \, .
\label{comconstr}
\end{equation}
The migratory (commuting) costs per trip for each edge $\cmig_{ij}$ ($\ccom_{ij}$) must be specified but are typically given as a function of distance with just one or two parameters. The total migratory (commuting) costs $B$ ($C$) are Lagrange multipliers which we will eliminate.

The model we have in mind is that we are looking for flows of emigration from site $i$ to site $j$ but the attractiveness of site $j$ depends on part on  the opportunities  at site $k$ provided the cost of commuting between $j$ and $k$ are reasonable, as illustrated in \fref{fig:commutingdemo}.
% \begin{figure}[htb]
%   \begin{center}
%   Sketch illustrating {\modelname}
%       %\includegraphics[width=30em]{movenearzipf}
%   \end{center}
%       \caption{A sketch showing interpretation of the commuting model.}
%       \label{fcommutingmodelsketch} %always put the label command after the caption for a figure or table.
% \end{figure}

The most likely pattern of flows is found by maximising this entropy subject to the constraints \eqref{outconstr}, \eqref{migconstr} and \eqref{comconstr}. This is done by finding the turning points in the function $\mathcal{L}$
\begin{eqnarray}
\mathcal{L}
   &=&
   \ln(T!) - \sum_{ijk} \left( T_{ijk} \ln\bigg(\frac{T_{ijk}}{N_{ijk}}\bigg) - T_{ijk} \right)
   + \sum_{i} \lambda_i \left( O_i - \sum_{jk} T_{ijk} \right)
   \nonumber
   \\
    &&
    \qquad \qquad
   + \beta \left( \Cmig - \sum_{ijk} T_{ijk} \cmig_{ij} \right)
   + \gamma \left( \Ccom - \sum_{ijk} T_{ijk} \ccom_{jk} \right) \, .
   \label{Lcaldef}
\end{eqnarray}
That is by demanding that $\partial\mathcal{L} / \partial T_{ijk}  = 0$ we find the most likely flow in this model which we find to be
\begin{equation}
T_{ijk} = N_{ijk} e^{-\lambda_i}  e^{-\gamma \ccom_{jk}} e^{-\beta \cmig_{ij}}  \, .
\end{equation}
We could impose the constraints on migration flows \eqref{migconstr} and on commuting flows \eqref{comconstr} but it is simpler to leave the expression in terms of the trip costs since they are usually expressed as a simple function of the distances between sites. However it is useful to impose the output constraint \eqref{outconstr} to eliminate the unknown Lagrange multipliers $\lambda_i$ and we find that we may rewrite the solution as
\begin{equation}
T_{ijk} = A_i O_i  e^{-\gamma \ccom_{jk}} e^{-\beta \cmig_{ij}}  \, .
\label{Tsolgen}
\end{equation}
where
\begin{equation}
A_i = \left[ \sum_{jk} N_{ijk}   e^{-\gamma \ccom_{jk}} e^{-\beta \cmig_{ij}} \right]^{-1} \, .
\end{equation}

At the moment we have left the choice for the number of microstates associated with each flow unspecified as $N_{ijk}$.  However in our work we choose
\begin{equation}
 N_{ijk} = Q_i Q_k \, ,
 \label{Ndef}
\end{equation}
where $Q_i$ is the population of site $i$. The reason is that the migration from one site to another will depend on the number of possible interactions between those two sites.  This is usually assumed to be proportional to the number of distinct possible trips.  That is we are assuming here that every person in site $i$ has $Q_k$ possible people to visit in site $k$ so the number of distinct trips, microstates in the language of statistical physics, is $Q_{i}Q_{k}$. In our situation, we are saying that the flow $T_{ijk}$ represents the number of people migrating from $i$ to site $j$ but that these people are working in site $k$.  We therefore argue that it is not the number of possible interactions between the origin site $i$ and the site of residence $j$ that matters, but it is the possible interactions between the origin site $i$ and the site $k$ with the opportunities proportional to $Q_k$.  We do not emigrate to Caldwell, Guildford or Suzhou because of the exciting interactions we will have in the sleepy suburb where we reside, but rather people emigrate  to Caldwell, Guildford or Suzhou because of the range of opportunities in the nearby site where they work: New York, London or Shanghai in these examples. Our definition for $N_{ijk}$ in \eqref{Ndef} is simple, though widely used in spatial interaction modelling, and it is for this reason that we have shown the more general solution for the flows in \eqref{Tsolgen}.

The other simplification, again a common one, is that we will set the output constraint $O_i=Q_i$ and along with our definition for $N_{ijk}$ in \eqref{Ndef} we arrive at the full form of the commuting model that we use in our work,
\begin{equation}
T_{ijk} = A_i Q_i  Q_k e^{-\gamma \ccom_{jk}} e^{-\beta \cmig_{ij}}  \, .
\label{Tsol}
\end{equation}
where
\begin{equation}
A_i = \left[ \sum_{jk} Q_k   e^{-\gamma \ccom_{jk}} e^{-\beta \cmig_{ij}} \right]^{-1}
\end{equation}

However, as we are updating the resident populations using the migratory movement, for two sites $i$ and $j$ this will just be $T_{ij} = \sum_{k} T_{ijk}$, yielding
\begin{equation}
T_{ij} = A_i O_i \left( \sum_k Q_k   e^{-\gamma \ccom_{jk}} \right) e^{-\beta \cmig_{ij}} \, .
\end{equation}
Writing this in the form of a traditional output-constrained gravity model, what we have is
\begin{eqnarray}
T_{ij} &=& A_i Q_i Q_j^\mathrm{eff} e^{-\beta \cmig_{ij}} \, ,
\\
A_i &=& \left[ \sum_{j} Q_j^\mathrm{eff}    e^{-\beta \cmig_{ij}} \right]^{-1}
\end{eqnarray}
where the attractiveness of site $j$, $D_j$, is the effective population $Q_j^\mathrm{eff}$ in the neighbourhood of the target site $j$
\begin{equation}
Q_j^\mathrm{eff}
 =
 \left( \sum_k Q_k   e^{-\gamma \ccom_{jk}} \right)
 =
 Q_j + \left( \sum_{k | k\neq j}  Q_k  e^{-\gamma \ccom_{jk}} \right)
\end{equation}
if we set $\ccom_{jj}=1$.

If this commuting model was used as a static model we see that the $Q_j^\mathrm{eff}$ are given in terms of existing input parameters $\{Q_k \}$ plus a second deterrence function $e^{-\gamma \ccom_{jk}}$.  It would be natural to express this second deterrence function in terms of the same set of known distances used for the $e^{-\beta \cmig_{ij}}$ deterrence function along with just one or two more parameters.  For instance we might set $e^{-\beta \cmig_{ij}}=e^{-\beta d_{ij}}$ and $e^{-\gamma \ccom_{ij}}=e^{-\gamma d_{ij}}$ using the same fixed set of distances $d_{ij}$. However even in this static case we see that we generate the measure for the output of each site, $O_i=Q_i$ is now different from the   ``attractiveness'' for input to a site, the $D_j=Q_j^\mathrm{eff}$ so even the static version of this model is distinctive.

This entropic Two-Trip framework is very flexible.  For instance the costs for either or both trips are just as easily expressed in terms of the Intervening Opportunities distance measure $s_{ij}$ of \eqref{intopp} \cite{S40}.  The Radiation model uses a cost function $\beta \cmig_{ij} = \ln\left((D_i+s_{ij})(D_i+s_{ij}+D_j)\right)$ with $\gamma \rightarrow \infty$ so that $D_i=Q_i$. However, if $\gamma$ is finite, then we find that we have $D_i=Q_i^\mathrm{eff}$ giving us a Two-Trip version of the Radiation model.

% **************************************************
\section{Discussion of the adjusted Ripley's K function}
\label{app:RipleyK}

The Ripley's $K$ function is used to detect deviations of a distribution of points from spatial homogeneity \citep{bevan2006multiscalar}. In our case (with a density of 1 in our units) Ripley's K function, denoted $K(r)$, is defined to be the average number of sites within distance $r$ of each site, that is
\tnote{Prefactor? Inverse density on Wikipedia but appendix suggest $1/n$ while our density is $1000/1000\times1000$.  In any case it is a constant prefactor so can be ignored.}
\begin{equation}
 K(r) = \frac{1}{\rho}\sum_{i \neq j} \theta(r-r_{ij})
 \label{RipleysK}
\end{equation}
where $\rho$ is the density of points, typically calculated as the number of points in a unit area of the system, $\rho = N/A$.

\begin{figure}[htb]
	\centering
    \includegraphics[width=0.9\textwidth]{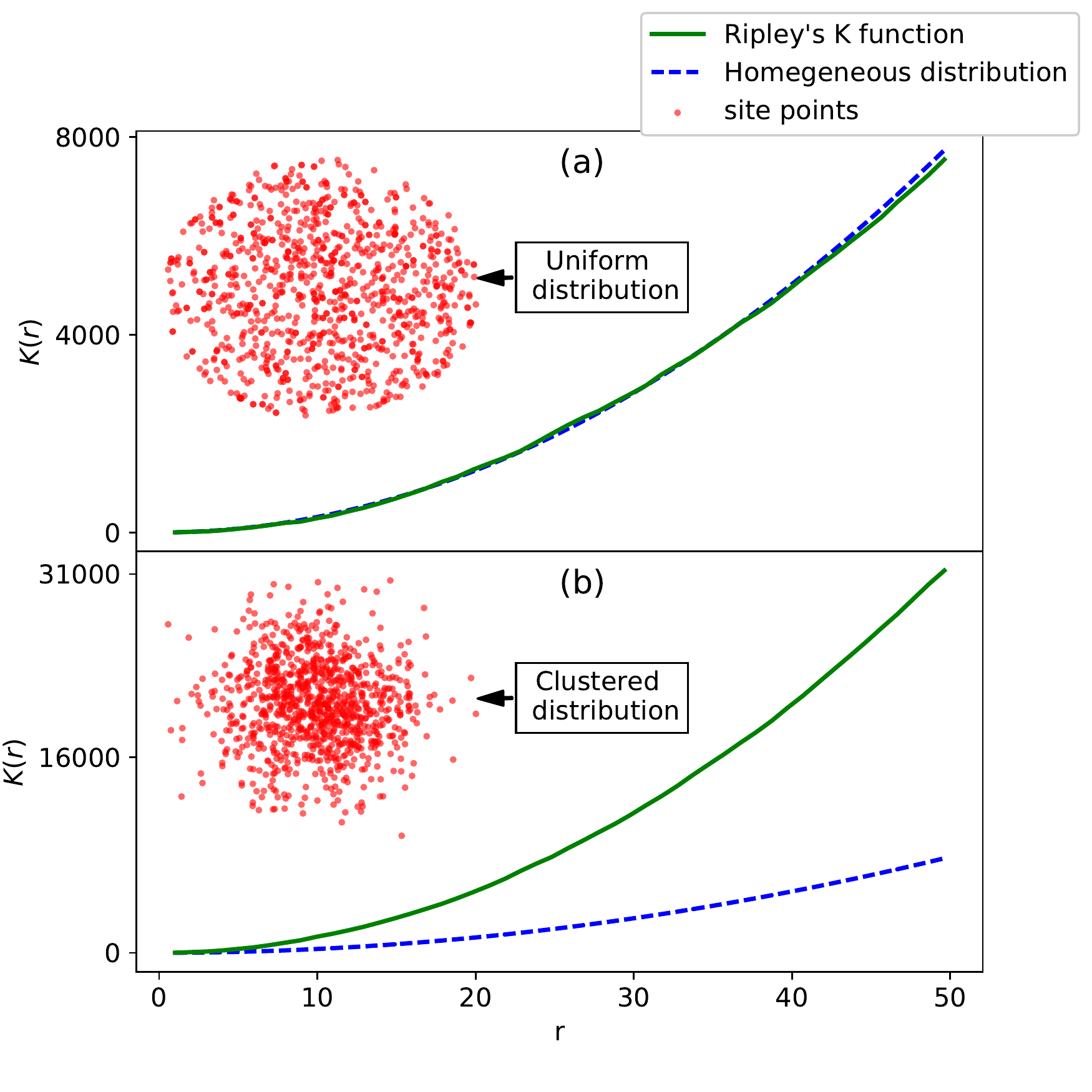}
    \caption{An illustration of the Ripley's K statistic on a set of randomly distributed points (a) and a set of clustered points (b).}
    \label{fig:ripleyseg}
\end{figure}

For points with perfect spatial homogeneity, Ripley's K function is of the form $K(r) = \pi r^2$. Any clustering of the points in space increases the value of the function. This behaviour is illustrated in \fref{fig:ripleyseg} %is this necessary?

Using the Ripley's K statistic presents two obstacles, both of which concern differences between the spatial characteristics of the data and the models. The first of these problems stems from ambiguity in the units of distance used in the models; and the second concerns the edge-effects arising from the shape of the area containing the settlements.

% --------------------------------------------------------------------------------------
\subsection{Distance Scales}
The first of these problems that we will address is the differing distance scales between our maps and United States census data. The result of this difference is that the two Ripley's K functions are incomparable due to two differing units of distance.
In order to overcome this, we introduce a scaling to distance units within the data, which effectively acts to normalise the density factors in equation \eqref{RipleysK}. We define the average distance within a system $\langle r \rangle$ as
\begin{equation}
\langle r \rangle = \frac{\sum_{ij} |r_{ij}|}{N(N-1)}
\end{equation}
where we have divided by $N(N-1)$ as this is the number of edges/connections present in the system. Then we scale the distance measured in data, $r_{d}$, to that of our models, $r_{m}$, as
\begin{equation}
r_{d} = r_{m} \times \frac{\langle r_{d}\rangle}{\langle r_{m} \rangle}
\end{equation}
So that both Ripley's K functions - both that of the model and that of the data - can now be calculated as functions of $r_{model}$ and comparisons between the two are now possible. Note, that this scaling must also be included within the definition of density, which must be transformed as
\begin{equation}
\rho_{d} = \frac{N}{\langle r_{d}\rangle^2} \times \left( \frac{\langle r_{d}\rangle}{\langle r_{m}\rangle} \right)^2
\end{equation}
where we have used the definition of map area $A = \langle r\rangle^2$ in calculating our density. Further to the above scaling, an additional scaling was used to place both $K(r)$ and $r$ parameters within the range $0 \to 1$, where 1 represents the maximum value measured in the models' outputs.

% --------------------------------------------------------------------------------------
\subsection{Edge Effects}

A second obstacle we face in using the Ripley's K statistic is the presence of edge effects. Whilst edge effects are the same (and therefore unimportant) between two identically shaped maps, the US has a far larger perimeter than our computationally generated square maps. Thus, we expect edge effects to have a higher order effect on the US census data than on the models - a feature that we must therefore account for. We can calculate the fractional change invoked in the Ripley's K function by a boundary by considering the Ripley's K statistic, but focusing on the total map \textit{area} within the boundary that falls within a radius $r$ instead of the total number of sites. This is equivalent to calculating the statistic on a uniform distribution of points within the boundary of the map. Thus we can rewrite the function as
\begin{equation}
 K(r) = \frac{1}{\rho}\sum_{x,y} A(r;x,y)
 \label{RipleysK2}
\end{equation}
where $x,y$ are uniformly distributed points within the map boundary and $A(r;x,y)$ is the map area that fall within radius $r$ of the point $\lbrace x,y \rbrace$ within the map. The significance of this measurement becomes clear when considering the result without a boundary and thus without edge effects. In this scenario, the function would \jnote{previously it said "on average" here. This isn't true, as the \textit{area} within the map always emulates a uniform distribution} simply take the form of $K(r) = \pi r^2$, and so we can take the ratio $R(r)$ between the map and its equivalent without a boundary.
\begin{equation}
  R(r) = \frac{\sum_{x,y} A(r;x,y)/n}{\pi r^2}
\end{equation}
Here $R(r)$ represents the fractional correction needed to account for edge effects in the Ripley's K function at radius $r$.

%% *****************************************************************
%\section{Other Material}
%
%% ......................................................................
%\subsubsection{Zipf and Population-rank distributions}\label{sec:zipfplots}
%
%The prominence of Zipf's law is illustrated for Brazil, the USA and India in \fref{fig:zipfs}.
%\begin{figure}[htb]
%  \centering
%  \includegraphics[scale = 0.8]{Zipfs}
%  \caption{Present day population-rank distributions for India, the United States of America and Braxil  for the largest one hundred (approx) settlements \citep{excel2014}. The distributions are plotted on logarithmic axes, so that Zipf's law is illustrated with a power-law fit (solid red line).}
%  \label{fig:zipfs}
%\end{figure}
%
%\tnote{I can not find this data explicitly on \citep{excel2014}. It appears you search and then you can find it as subpages.  I suggest place the data in our own figshare.com and there we add links to the sources.  For instance the USA data I found appears to come from the Wikipedia page \url{https://en.wikipedia.org/wiki/List_of_United_States_cities_by_population}.}
%
\end{document}